\documentclass[lettersize,journal]{IEEEtran}
\usepackage{amsmath,amsfonts}
\usepackage{algorithmic}
\usepackage{algorithm}
\usepackage{array}
\usepackage[caption=false,font=normalsize,labelfont=sf,textfont=sf]{subfig}
\usepackage{textcomp}
\usepackage{stfloats}
\usepackage{url}
\usepackage{verbatim}
\usepackage{graphicx}
\usepackage{cite}
\begin{document}

\title{An Empathetic User-Centric Chatbot for Emotional Support}

\author{ Yanting Pan, Yixuan Tang, Yuchen Niu

}

\maketitle

\begin{abstract}

This paper explores the intersection of Otome Culture \cite{b1}and artificial intelligence, particularly focusing on how Otome-oriented games fulfill the emotional needs of young women. These games, which are deeply rooted in a subcultural understanding of love, provide players with feelings of satisfaction, companionship, and protection through carefully crafted narrative structures and character development. With the proliferation of Large Language Models (LLMs), there is an opportunity to transcend traditional static game narratives and create dynamic, emotionally responsive interactions. We present a case study of Tears of Themis,  where we have integrated LLM technology to enhance the interactive experience. Our approach involves augmenting existing game narratives with a Question and Answer (QA) system, enriched through data augmentation and emotional enhancement techniques, resulting in a chatbot that offers realistic and supportive companionship. This paper also proposes a set of generic methodologies for improving the training of role-playing artificial intelligence, aiming to set a new standard in interactive entertainment and emotional engagement within digital spaces.

\end{abstract}


\section{Introduction}

The evolution of chatbots has been marked by significant milestones, transitioning from simple rule-based systems to advanced, Large Language Models (LLMs) driven platforms. These modern chatbots, leveraging AI like GPT-4, exhibit unparalleled linguistic capabilities, understanding context and generating human-like responses. 

In our project, we've merged new data augmentation method to augment the quality and diversity of role-playing data.

\section{Background}
Using the chapter Tears of Themis\cite{b2} as a case study for Otome games, this paper delves into the pivotal role of the male protagonist's character design and image in game development. The essence of the character, beyond the visuals crafted by artists and the accompanying voice acting, is primarily conveyed through the narrative. The dialogue and actions of the male leads within the game's storyline enrich their characterizations, forming shared memories with the player through interaction. Moreover, to maintain player engagement and support long-term operations, "Tears of Themis" periodically introduces new events, developing different cards for the male leads that supplement the storyline and reinforce their characterizations, thereby enhancing player interaction.
"Tears of Themis" features four male protagonists, each with a distinct personality. The success in character development is evident in the balanced popularity among the characters and the vividness of their traits, with narrative progression and supplements remaining consistent with their established characterizations. The fundamental allure of Otome games lies in the deep emotional experiences and bonds formed with virtual characters. However, the operational and human resource limitations of gaming companies impede frequent updates to storylines and interactions, leading to a perceived lack of interactivity and an inability to fully satisfy players' emotional needs. Given the comprehensive data and well-rounded characterizations in Otome games, along with a strong demand for emotional companionship, this paper proposes a novel LLM-based, emotionally supportive role-playing bot. With the four male leads from "Tears of Themis" serving as a demo, the training methods outlined here can be generalized to any role-playing bot that focuses on personality and emotional connection.

\section{Related Work}
While LLMs excel in information processing and conversation\cite{b3} \cite{b4} \cite{b5} , their application in providing emotional companionship, particularly in simulating characters from emotional games, remains underexplored. There's a noticeable gap in the domain of emotionally supportive bots. This presents a unique opportunity for innovation in creating chatbots that offer more than factual assistance, venturing into the realm of emotional intelligence and empathy.

The current landscape of psychotherapeutic bots, such as Wysa\cite{b6}, represents a significant stride in mental health support. These bots employ cognitive-behavioral techniques and guided conversations to aid users. However, their approach often borders on being formulaic, lacking the nuanced understanding of human emotions and personalized responses. This limitation highlights the need for a more adaptive and empathetic approach in therapeutic bots, one that can tailor conversations and responses based on individual emotional states and needs. The challenge lies in imbuing these bots with a deeper sense of emotional intelligence, enabling them to navigate the complexities of human emotions more effectively.


\section{Conversation Design}
The dialogue between the bot and users permeates the entire interaction, making the phrases generated for communication essential.

Our chatbot requirements could be divided into two aspects. On one hand, there is a need for objective knowledge related to the game. This knowledge is required to provide accurate responses in conversations with any character, regardless of who the user is interacting with. For example, (insert specific examples here).

On the other hand, we have the casual conversation aspect with the user. In this part, the questions may not necessarily be related to the game, but they must align with the personalities of different characters, including tone and other characteristics. Furthermore, the conversations should adhere to standard conversational norms, such as coherence.

\section{Data collection and augmentation}
With the rapid advancement of large models such as ChatGPT, continuous tests have revealed that the quality of data often plays a pivotal role in determining the effectiveness of a language model. High-quality data can substantially boost the accuracy of such models, emphasizing the critical importance of creating a robust dataset.

\subsection{Initial Data \& Preliminary Augmentation}
{\bf Basic Game-related Questions:} 
For Task One, we conducted data scraping on the official website and relevant forums related to the game. This primarily covered objective information in four aspects: the background of the cases, facts related to the cases, individuals involved, and the truth behind the cases.

For instance:
\textit{\{"prompt": "Do you still remember what happened to reporter Kong Moli?","answer":"Of course, journalist Kong Molli was sued by the Haiaosen Group for damaging their business reputation after she exposed their severe pollution. Wang Han and Qiu Heng, who were bribed, murdered Kong while she was gathering evidence and staged it as a car accident. It's truly a tragedy, but I believe that the wicked will pay for their greed."\}}

For this data, we enlisted the help of four experienced gamers to manually generate potential questions based on this information and provide correct answers. Players were allowed to add more detailed questions while ensuring the accuracy of the answers.

From this, we gathered 80 genuine sentences for base questions training. We manually screened and filtered out data that didn't meet our criteria (too short, off-topic, inconsistent with correct answers).


{\bf Casual Conversations Related to Character Personas:} Initially, based on the chatbot's flowchart, we formulated 40 causal questions. Here is an example of a dialogue with responses from four characters:\\
\textit{\{"prompt": "If you could choose a place to travel, where would you most like to go?"\}}\\
\textit{\{"answer by character1 - Xia Yan":"You know, what I really want is to go to Yunnan with you, to visit the Cherry Blossom Valley, Puji Temple, and the ancient Tea Horse Road! I've heard about the Wa people's custom of 'declaring love through leaf whistling.' It's said that when the green leaves whistle, if birds flutter and fly around, the love will last forever. I hope to have the chance to whistle it for you."\}}\\
\textit{\{"answer by character2 - Zuo Ran":"I don't have a specific place I long to visit; I simply want to accompany you to every place you cherish. I'll plan everything in advance to ensure you enjoy our travels to the fullest."\}}\\
\textit{\{"answer by character3 - Mo Yi":"If you're asking about a place... well, there is one. As someone who loves wine tasting, I've been wanting to visit Bordeaux. It's home to many vineyards, and in the fall, you can experience grape picking and wine making. We could travel by train and enjoy the scenic journey. Does that interest you?"
\}}\\
\textit{\{"answer by character4 - Lu Jinghe":" Among the places I've visited, Florence holds the crown as my favorite city in Europe. Words fail to capture its romantic essence. After art classes at school, I used to sit in Michelangelo Square, watching the sunset, finding inspiration in this city. It healed many of my nights, and now, you are my remedy. I hope for a chance to take you to Florence again."\}}

We made the four experienced players to write appropriate responses based on the personalities and tones of these four character personas. We encouraged players to incorporate relevant story details into their responses as well. In the end, each character persona had a set of paired Q\&A dialogues.

We first employed the traditional data augmentation method, MarianMT’s Back Translation\cite{b16}, where we first translate the questions into another language (like German) and then translate them back to the original language. Using this approach, we could expandthe initial 40 sentences in the Casual Conversations task to 1,000 sentences.


\subsection{Self-instruct}
However, the above-mentioned method for data generation is inefficient. Through testing, 60.2\% of short sentences had a ROUGE-L score higher than 0.7. We've opted to apply the self-instruct method for data augmentation based on the open source language model LLaMA-7b.

The self-Instruct\cite{b39} approach is a partially automated iterative bootstrapping process that utilizes a pretrained LLM's capabilities to produce a diverse set of instructions. It begins with initial manual instructions and then iteratively generates new tasks and enriches the task set by removing low-quality or redundant instructions.

This data augmentation process needs to be used in conjunction with a prompt\cite{b44}. 
\begin{itemize}
\item From our experiments, the most crucial aspect regarding instructions is clarity and specificity (e.g., “Provide three rephrasings for the following question").
\item Next, enriching the context and establishing a persona could increase the diversity of output results. The model could also simulate a persona directly by describing specific characters.
\item Lastly, additional requirements could be incorporated (e.g., "Use clear and concise language and write in a confident yet friendly tone; make it conversational," and so on).
\end{itemize}

Through self-instruct, we've augmented the original data by threefold: 240 in Basic Game-related Questions and 3,000 in Casual Conversations Task.

\begin{figure}[h!]
\centering
\includegraphics[width=0.5\textwidth]{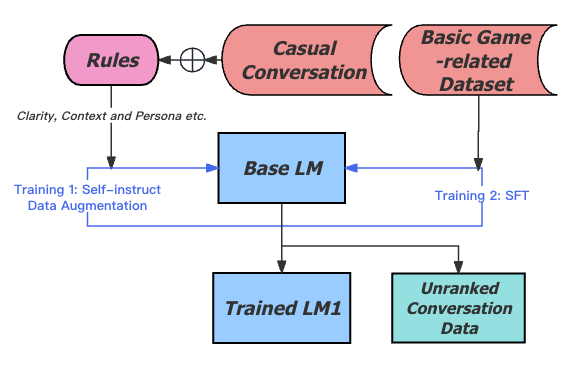}
\caption{\label{fig:sft}Self-instruct \& SFT training}
\end{figure}

\section{Implementation}

The current prominent Language Model (LM) implementations, such as ChatGLM\cite{b47} and Chinese-Alpaca\cite{b48}, typically involve three essential steps: 
1. Knowledge Acquisition and Pre-training:
Initial model training on extensive data for understanding context.
2. Instruction-based Fine-tuning:
Fine-tuning the model based on specific instructions and optimizing memory usage.
3. Preference Refinement:
Enhancing model outputs to align better with human preferences using Reinforcement Learning from Human Feedback (RLHF)\cite{b44} or Differentiable Programming Optimization (DPO)\cite{b43}.

We choose LLaMA 2\cite{b38} as our base model. This choice stems from its promising performance, as demonstrated in benchmark tests where it consistently outperforms ChatGPT, particularly in generating helpful prompts\cite{b37}. Moreover, being an open-source model, it provides us with the freedom to tailor the model to our specific requirements, in our case, developing our own chatbot.


\subsection{SFT}
We conduct an initial fine-tuning using Ultrachat dataset\cite{b7} (774k for training), a second fine-tuning is executed utilizing our own Basic Game-related Questions dataset. 

The loss for this task is calculated by computing the cross-entropy between the answers generated by the model $U_{Model}$ and our reference answers $U_{Reference}$. 
\begin{equation}
LEC =CrossEntropyLoss(U_{origin},U_{rewriting})
\end{equation}

Each SFT sample consists of two parts: prompt + answer, such as:

\textit{\{"prompt": "How many administrative districts does Weiming City have?","answer":"There are four zones in total. They are Jianan District, Binhe District, Jinlan District and Changtan District."\}}

\subsection{Human Preference Data Collection}
In order to enhance the quality of conversations, we aim to leverage the data collected earlier by training the language model to better distinguish between good and bad sentences, aligning its language with various character personas, and ultimately achieving smoother dialogue. Thus, collecting preference data is unavoidable. The results of human feedback are regarded by us as the ultimate judgment criteria. Yet, relying solely on manual annotation is too time-consuming. Therefore, we first manually annotate a batch of reliable high-quality data, and then use this annotated data to train a reward model. Subsequently, we utilize this reward model to label the remaining unannotated preference data. Finally, we employ GPT-3 for scoring\cite{b45} \cite{b46}, comparing its scores with the scores generated by our trained reward model to reevaluate the quality of the data annotations.

\begin{figure}[h!]
\centering
\includegraphics[width=0.5\textwidth]{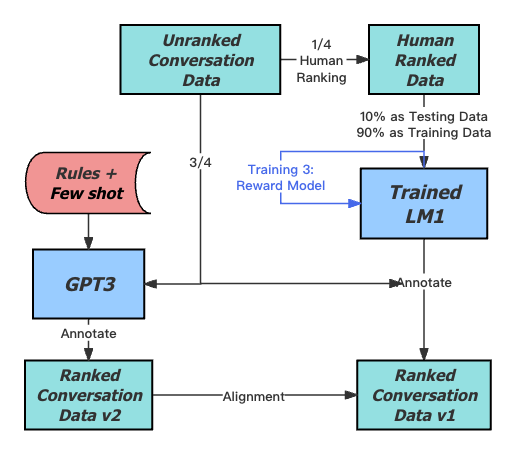}
\caption{\label{fig:rm}Reward Model Training}
\end{figure}

{\bf Training \& Testing Data }- Human Feedback: In order to avoid bias, we invited three annotators to participate in the annotation process. For the original 1,000 empathy rewriting data, since our paraphrase for each seed sentence is more than 10 per sentence, it would be difficult to apply pair-wised comparison method. Thus we did not use the popular pair ranking method for sorting but had annotators assign scores of 0(non-empathetic), 1(mild empathy), or 2(strong empathy). The final result is determined by the score that receives the greatest number of votes. Sentences with split scores underwent further assessment. Through the combination of high scores and low scores from the same seed, we finally got 10742 pairs of Reward Model data, from which we extracted 1000 as the test set, and the rest 9742 as training set.

{\bf Reward Model}: Training the Reward Model involves a sorting task where chosen and rejected responses are inputted for a given query $x$. After computing embedding for each token, they are fed into a linear layer to generate scores for each token. The score associated with the final token serves as the reward score $r_{\theta}$.

The objective during training is to maximize the distinction between chosen $y_{w}$ and rejected $y_{l}$ responses. This is achieved by optimizing a loss function\cite{b40}:

\begin{equation}
loss\left(\theta\right)=
-{E}_{(x,y_{w},y_{l})\sim D}\left[\log\left(\sigma(r_{\theta}\left(x,y_{w}\right)-r_{\theta}\left(x,y_{l}\right)\right))\right]
\end{equation}
where $\sigma$ is the logistic function. During the training process, each batch may contain different prompts. The loss is divided among all batches.

{\bf Training}: We controlled training to 1 epoch\cite{b51} to avoid overfitting\cite{b50}. Using self-instruct, we generated three paraphrases for each of the original 1,000 sentences(3000 in total). We then submitted them to our trained reward model for scoring. 

{\bf Alignment} - GPT3 labeling vs Reward Model labeling: We finally selected and submitted 100 pairs for evaluation using GPT-3. We established four criteria for evaluation: empathy in the reply, appropriateness of tone, redundancy, and similarity in meaning to the original sentence\cite{b41}. When constructing the prompt, we include the names and character settings of different personas in our system prompt in order to increase evaluation accuracy. The results were then sorted based on them. The outcomes revealed a 76\% alignment with the scores generated by our reward model, affirming the reliability and consistency between our reward model and GPT-3 scoring\cite{b42}.

Finally, we got 13753 ranked data pairs in total for the following process.

\subsection{DPO}

RLHF is time-consuming and labor-intensive. In addition to the Reward Model, during the PPO stage, it requires loading four models - two for inference and two for training. In contrast, DPO is relatively straightforward, requiring only two models to be loaded, one for inference and the other for training. Training could be performed directly on preference data, and the results are satisfactory. Therefore, we choose to adopt the DPO algorithm.

{\bf Loss}:
\begin{equation}
\resizebox{.98\hsize}{!}{${\mathcal{L}}_{\mathrm{DPO}}(\pi_{\theta};\pi_{\mathrm{ref}})=-\mathbb{E}_{(x,y_{w},y_{l})\sim D}\left[\log\sigma\left(\beta\log{\frac{\pi_{\theta}(y_{w}\mid x)}{\pi_{\mathrm{ref}}(y_{w}\mid x)}}-\beta\log{\frac{\pi_{\theta}(y_{l}\mid x)}{\pi_{\mathrm{ref}}(y_{l}\mid x)}}\right)\right]$}
\end{equation}

$\pi_{ref}$ is initialized from the SFT model $\pi^{SFT}$, $\pi_{\theta}$ is the language model policy that we want to optimize, $\beta $ is a parameter controlling the deviation from the base reference policy $\pi_{r e f}$.

The key is higher rewards for chosen over rejected responses (model favors chosen responses).

\section{Evaluation }
Following the dual fine-tuning process\cite{b49} , the LLaMA2 model achieved an accuracy of 0.8104 and a Macro-F1 score of 0.8195. These metrics demonstrated a notable enhancement compared to the baseline, with a 60.67\% increase in accuracy and a 67.24\% improvement in Macro-F1 score. The baseline is obtained by having three individuals who have never played the game answer our sampled 20 questions and then compile the results for analysis. Consequently, this refined model was chosen for the following tasks.
\begin{table}[h!]
\centering
  \begin{tabular}{|c|c|c|}
 \hline
 Model & Accuracy & Macro-f1 \\ [0.5ex] 
 \hline
 \hline 
 Baseline & 0.5044 & 0.4900  \\ 
 LLaMA 2 & 0.5860 & 0.5183 \\
 \textbf{Otome bot} & \textbf{0.8104} & \textbf{0.8195} \\
 \hline
 \end{tabular}
 \caption{Basic Game-related QA Accuracy}
\label{table:emoclass}
\end{table}

\begin{table}[h!]
\centering
  \begin{tabular}{|c|c|c|}
 \hline
 Model & Rouge-L  \\ [0.5ex] 
 \hline
 \hline 
 LLaMA 2 & 0.1180 \\ 
 \textbf{Otome bot Character1}  & \textbf{0.5195} \\
 \textbf{Otome bot Character2}  & \textbf{0.4988} \\
 \textbf{Otome bot Character3}  & \textbf{0.5009} \\
 \textbf{Otome bot Character4}  & \textbf{0.4824} \\
 \hline
 \end{tabular}
 \caption{Evaluation of our model using Rouge-L.}
\label{table:emoclass}
\end{table}

Our model has also demonstrated significantly notable achievements in its capability to mimic four characters, exhibiting substantial improvements over the baseline. 


\section{Limitations}
Our project, while pioneering in its approach, faces several limitations. Firstly, our data selection is confined to just a single chapter of the game, leaving a vast array of game characters and narratives untapped. This indicates a significant potential for expanding our database to encompass a more comprehensive range of game content, thereby enhancing the chatbot's versatility and depth in character portrayal.

Additionally, the advent of multi-modal interactions supported by GPT-4 opens new avenues for improvement. Our current chatbot lacks this multi-modality, relying solely on text-based interactions. Integrating visual, auditory, or other sensory data could dramatically enrich the user experience, making interactions more immersive and engaging.

In terms of user interface design, there is room for aesthetic enhancement. A more visually appealing and user-friendly interface could significantly improve user engagement and satisfaction, making the chatbot more accessible and enjoyable to interact with.

Finally, our chatbot exhibits a degree of forgetfulness in conversations. It struggles to retain and recall past dialogues, which can impede the flow of interaction and the building of a coherent, continuous character narrative. Addressing this issue by improving memory retention capabilities would make the chatbot more realistic and effective in sustaining engaging conversations.

\subsection{Conclusion}

This project marks as the first chatbot developed in conjunction with this specific Otome
Culture game. Our chatbot breaks new ground by adopting a more relaxed and youthful approach, particularly addressing a gap in chatbots focused on women's issues, instead of a serious or didactic tone. This represents a substantial advancement in making mental health support more accessible and engaging for a female audience, a demographic often underrepresented in this technology domain.

In terms of data augmentation, we have innovated beyond the basic self-instruct approach, aligning more closely with human-like interactions. This has not only enhanced the authenticity of our chatbot's responses but also improved its ability to empathize and engage with users on a more personal level.

While acknowledging the limitations of our current project, including the need for a broader database, multi-modal interaction capabilities, improved UI design, and better memory retention, these areas also outline our future direction. We are committed to evolving our chatbot, expanding its capabilities, and refining its interactions. Our aim is to continually advance the field of empathetic chatbots, specifically tailored for emotional support and companionship, especially for women. By pushing the boundaries of current chatbot technology, we hope to create a more inclusive and emotionally intelligent digital landscape.


\vspace{12pt}

\end{document}